\def\be{\begin{equation}}
\def\ee{\end{equation}}
\def\ber{\begin{eqnarray}}
\def\eer{\end{eqnarray}}
\def\bers{\begin{eqnarray*}}
\def\eers{\end{eqnarray*}}

\def\JPC{J. Phys. C}

\def\PR{{ Phys. Rev.}\ }

\def\JPC{{ J. Phys. C: Solid State Phys}\ }

\documentclass[aps,prb,twocolumn,groupedaddress,showpacs,amsmath,amssymb]{revtex4}
\usepackage{dcolumn}   
\usepackage{amsmath}
\usepackage{graphicx}    
\usepackage{subfigure}
\usepackage{color}

\newcommand{\comment}[1]{}
\usepackage{ifthen}
\newboolean{includefigs}
\setboolean{includefigs}{true}      
\newboolean{includetext}
\setboolean{includetext}{true}      
\newcommand{\condcomment}[2]{\ifthenelse{#1}{#2}{}}

\begin{document}

\title{Efficient Isoparametric Integration over Arbitrary, Space-Filling Voronoi Polyhedra for Electronic-Structure Calculations}

\author{Aftab Alam,$^{1,3}$ S. N. Khan,$^{2,3}$, Brian G. Wilson,$^{4}$, and D. D. Johnson$^{1,2,3}$}
\email[email: ]{ddj@ameslab.gov}
\affiliation{$^1$Department of Materials Science and Engineering, University of Illinois, Urbana, IL 61801, USA }
\affiliation{$^2$Department of Physics, University of Illinois, Urbana, IL 61801, USA}
\affiliation{$^3$Division of Materials Science and Engineering, Ames Laboratory, Ames, Iowa 50011, USA }
\affiliation{$^4$Lawrence Livermore National Laboratory, Livermore, California 94550, USA }

\begin{abstract}
A numerically efficient, accurate, and easily implemented integration scheme over convex Voronoi polyhedra (VP) is presented for use in {\it ab-initio} electronic-structure calculations. 
We combine a weighted Voronoi tessellation with isoparametric integration via Gauss-Legendre quadratures to provide rapidly convergent VP integrals for a variety of integrands, including those with a Coulomb singularity.
We showcase the capability of our approach by first applying to an analytic charge-density model achieving machine-precision accuracy with expected convergence properties in milliseconds.
For contrast, we compare our results to those using shape-functions and show our approach is greater than $10^{5}$  faster and $10^{7}$ more accurate.
A weighted Voronoi tessellation also allows for a physics-based partitioning of space that guarantees convex, space-filling VP while reflecting accurate atomic size and site charges, as we show within KKR methods applied to Fe-Pd alloys.
\end{abstract} 
\date{\today}
\pacs{02.60.Jh, 71.15.Dx  }
\maketitle


{\par}A variety of  science and engineering research problems require multicentered integrals that cannot be solved analytically due to the complex domains of integration. 
The total energy and potential in any site-centered, electronic-structure calculation involves the evaluation of three-dimensional integrals over convex Voronoi polyhedra (VP).\cite{Voronoi08} 
 Although a number of numerical integration techniques have been proposed,\cite{Ellis70,Boerr88,Averil89,Velde92,Finocchiaro98} an efficient, accurate, reliable and easily implemented scheme is still lacking. 
Prior methods often rely on detailed analysis of symmetry properties of the integration domains and, hence, limit their applicability to arbitrary atomic geometries and structures.  
A major continuing need is an integration method over space-filling VP that has a high degree of accuracy with a minimal computational effort and that is sufficiently generic so that it can be used in most electronic-structure application codes.
 
{\par}  For example, ``exact'' linear muffin-tin orbital (EMTO) method\cite{EMTO} uses an approach from Gonis \emph{et al.}\cite{Gonis1991} to overcome VP integration issues for the Poisson potential, but it is extremely slowly convergent; various KKR-based codes, such as the linear-scaling multiple-scattering (LSMS),\cite{LSMS} utilizes shape-functions to perform VP integrations, which, as we show, is slowly convergent and limited in accuracy; the  full-potential linear augmented wave\cite{FLAPW} (FLAPW) method avoids VP integrals (via non-overlapping muffin-tins and Fourier methods over entire unit cell), but never determining site-VP-specific properties and requiring a larger number of spherical harmonic basis functions and huge number of plane-waves.

{\par} We present such an algorithm by combining a weighted VP tessellation\cite{Voronoi08,Tanemura-83}  with isoparametric methods\cite{Isoparametric} to provide rapidly convergent integrals for various integrands, including Coulomb singularities.
For generality, we use a Radical Plane Construction\cite{Gellatly81} (RPC) or Power Diagrams\cite{Aurenhammer} to guarantee convex, space-filling VP.
For electronic-structure use, physics-based weights are optimally chosen as ratios of radii determined from the topology of the electronic density.\cite{Alam09}
Isoparametric transformations then permits analytical mapping of polyhedra subdomains to a bi-unit cube, which are then simply integrated by Gauss-Legendre quadratures. 
For any VP we only need evaluate numerically the integrands, Jacobian and weights at Gauss points for a relatively fast and accurate integral, and no issues with divergence.

{\par} We showcase our isoparametric method by two  means.
First, we evaluate various electronic integrals analytically using a well-known charge density model by van Morgan,\cite{Morgan77} and show directly the accuracy and efficacy of our numerical method.
Second, we implement the method in an all-electron Korringa-Kohn-Rostoker (KKR) code\cite{MECCA}  and apply it to a phase stability study of face-centered-cubic,  (dis)ordered FePd. We exemplify the accuracy for formation enthalpies and the insensitivity of results to the choice of spherical-harmonic basis-set ($L$-expansion) due to the use of weighted VP. 

{\par}We organize the paper as follows: After Section I background, we describe in Sec.~II the RPC tessellation and weights that we merge with an isoparametric integration via an analytic dual-coordinate transformations, known in the finite-element community, to create a general and optimal integration scheme.
In Sec. III, we describe the van Morgan charge-density model  to assess the performance of any integration  method. 
In Sec.~IV, we address the accuracy, convergence, and timings of this isoparametric scheme for  close-packed structures; machine-precision accuracy with expected convergence is found, with millisecond timing for each VP. 
Our approach is greater than $10^{5}$  faster and $10^{7}$ more accurate than that with shape-functions. 
Finally, we discussed the results for application to FePd, then conclude in  Sec. V.

\section{Background}

\begin{figure}[t]
\centering
\includegraphics[width=7.5cm]{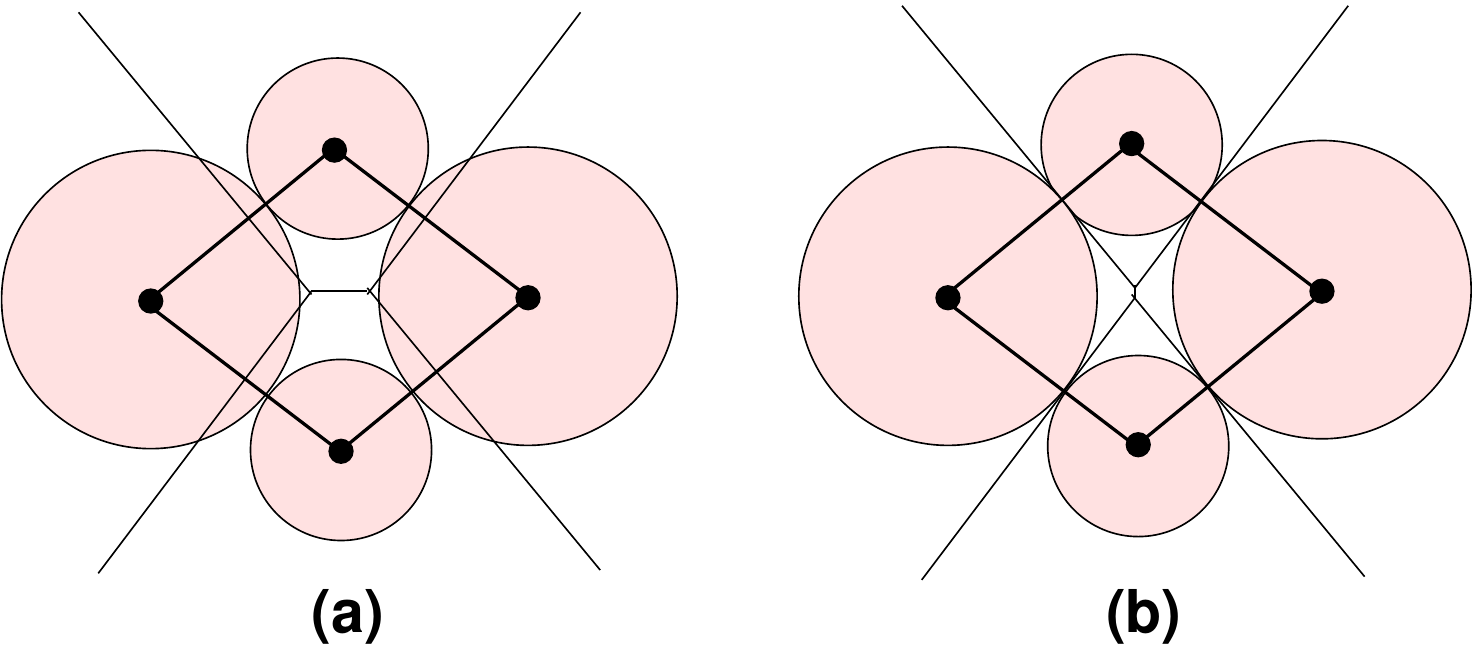}
\caption
{(Color online) Tessellation via (a) geometry and (b) RPC. The ``size'' of little (big)  atoms is over (under) estimated in (a), affecting properties, whereas (b) reflects atomic ``size'' if radii are determined from charge densities, see text.}
\label{weight_voronoi}
\end{figure}

{\par}Given any atomic configuration, the first step to compute any site-centered integral quantity is to perform a Voronoi tessellation of the space in and around the molecule or solid, including possible \emph {empty sites} to improve, for example, a site-centered basis set. 
Standard geometric (i.e., the Wigner-Seitz) tessellation  subdivides space into VP such that every point within a VP cell has the property of being closest to one and the same site, and that each polyhedral face is orthogonal to and bisects the line segments joining the site centers, see Fig.~\ref{weight_voronoi}.
However, in most materials (e.g., size-mismatched alloys), it is a poor subdivision of space,  making VP corresponding to the smaller (larger) atoms too large (small),\cite{Alam09} see Fig.~\ref{weight_voronoi}.
 Such errors impact solid-state and biological problems, where, e.g., an accurate interstitial volume is required for reliable predictions of thermostability of cavity-filling mutants in proteins,\cite{Goede97} or for statistical models in continuum systems where packing geometry plays a key role.\cite{Sastry97}

{\par}Given these deficiencies, various proposals have been made to place the dividing plane subject to atomic size. 
 Richard\cite{Richard74} suggested using the ratio of the distance between atoms and dividing plane to be equal to the ratio of the corresponding atomic radii, but this does not reflect bonding.
RPC weights the distance to each atom by subtracting the squared atomic radius from the squared Euclidean distance,\cite{Gellatly81,Aurenhammer} which guarantees convex and space-filling VP, but radii must be provided.
Other generalizations include the introduction of non-planar boundaries between atoms.\cite{Medvedev-Gerstein}  
For site-centered methods, the convex, space-filling property is critical; for example, in KKR the scattering matrices are only defined for convex VP, but the spherical-harmonic basis must reflect accurately the spherical density of each site or else the basis must be augmented (e.g., plane waves in FLAPW,\cite{FLAPW} or spherical waves\cite{ASW-ref}).
Notably, the VP tessellation benefits from a judicious choice of weights related to electron density.\cite{Alam09}

{\par} With a Voronoi tessellation in hand, a scheme must be chosen to perform VP integrations. 
In one-dimension, there are many numerical  techniques that easily achieve accurate results. 
 Yet, accurate and fast techniques for three-dimensional integrands with unusual domains, found in molecular or solid-state calculations, remain an active area of study.\cite{Xiao} 
 We mention some key previous work. 
Ellis and Painter\cite{Ellis70} used Diophantine integration\cite{Hasel-Conroy} in molecular calculations, with convergence as $O(N^{-1/2})$ to $O(N^{-1})$ for N sample points, just better than Monte Carlo. 
Becke \cite{Becke88} used standard Voronoi partitioning with simplifications introduced to reduce multicenter integrals to a sum of single-center ones, with less detailed convergence studies than we provide here;  we find that it uses slightly more points than ours for low ($\sim$$10^{-3}$) accuracy and much slower convergence for higher precision.  
More recently, Gaussian product formulas have been found useful when awkward domains of integration were split into tractable subdomains.\cite{REF-GaussProduct}
It is, however, not only the form of the integration domain but also behavior of the integrand that may necessitate the  use of product formulas and further subdivision of the subdomains. 

{\par} For site-centered basis sets, there are limited choices of  subdivision for convex VP domains.
Baerends and coworkers\cite{Boerr88,Velde92} broke each VP into subpolyhedra formed by the nucleus and its base of one of the VP faces followed by  further subdivision of the face into connected sets of triangles and quadrilaterals. 
Averill and Painter\cite{Averil89} cropped each VP by an inscribed sphere to form an interstitial region associated with each VP face; hence, interstitial integrals are expressed as a sum of integrals over cropped pyramids.
(In finite systems, a separate subdomain is the bounding part of the space outside the local atomic VP.)
We use an analytic transformation to a bi-unit cube for the cropped integrals, an approach similar, but not the same as, Baerends and coworkers,\cite{Boerr88,Velde92} as we discuss.
Also, previous methods never considered using the underlying charge density to chose an optimal subdomain of integration.
In  all-electron methods,  the wavefunction and potential have cusps and singularities near the nuclei, respectively; these functions are easily integrated over spherical domains, although the ``interstitial'' region (between the sphere and VP facet) has a complex shape. 
Uniquely, we utilize the charge-density topology and the behavior of the integrands to determine the VP (spherical and interstitial) subdomains.

{\par} Although various integration methods have been proposed, the accuracy attained has usually been poor compared to the computational effort expended, typically a modest ($7-8$ digit) accuracy required a large number of sampling points. 
The desire, of course, is to approach machine-precision accuracy with a modest number of sampling points. 
While noting some similarities with previous methods,\cite{Averil89,Velde92} the present approach is unique in features and, particularly, its efficiency and accuracy for polyatomic systems; also it has the advantage of being conceptually simple and easy to implement.
We verify that accurate ($14$ digit) integration over various kernels is achieved with a modest number of Gauss points.
Moreover, when combined with the use of physics-based weighted VP, an insensitivity to site-centered basis set is possible while achieving high accuracy, as we show.

\section{METHOD} \label{method}
{\par} The partitioning of space in and around a molecule or solid into convex polyhedra by RPC is described, followed by an analytic dual-coordinate transformation (isoparametric mapping) of a bi-unit cube to obtain the shape of any specific facet subdomain of each VP.
Dissection of each VP can be accomplished in two ways: Either each VP  (1) is divided into subdomains formed by the pyramid between face and a sites origin; or (2) is split into an inscribed sphere domain and a sum of interstitial domains between sphere and each facet plane. 
We then find the integral at each VP as a sum over the Gauss quadratures by numerical evaluation of the integrands, weights and Jacobian of the transformation. If the integrand is evaluated at known Gauss points, only sampling points determine the error; whereas, if a numerical (discrete) grid is used, then interpolation error needs to be ameliorated.

\subsection{Weighted Voronoi Radical Plane Construction} 
{\par} In unweighted tessellations, the subspace of points closer to an atom centered at ${\bf r}_i$ than any other atom centered at ${\bf r}_j$ is considered the VP about atom ${\bf r}_i$. 
For an atom on the edge of a molecule, the ``polyhedron" will be unbounded; that can be amended by introducing an extra dividing plane that is tangent to the inscribed sphere of the unbounded cell. 
The best choice will be the plane that also minimizes the volume in the resulting cell. 
Of course, no such edge difficulties can occur for periodic crystals. 
In the case of a one-atom crystal, the VP is simply the Wigner-Seitz (geometric) cell.

{\par} For a crystal with atoms of unlike size, the standard Voronoi partitioning of space assigns too much volume to the smaller atoms, see Fig.~\ref{weight_voronoi}(a), and the corresponding  VP is unphysical.\cite{Alam09} 
The biased partitioning incorrectly determines the atomic  volumes and the VP structure. 
A weighted tessellation corrects  by re-weighting the distances to the site centers. 
This re-sizing allows the volume given to  each site to grow or shrink in accord with its size (radius). 
RPC uses the weighted metric $\vert\vert {\bf r}-{\bf r}_{i} \vert\vert^2 - R_{i}^{2}$, where $R_i$ is the radius of the $i^{th}$ atom and $\vert\vert {\bf r} - {\bf r}_i \vert\vert$ is the Euclidean distance between point ${\bf r}$ and atom center ${\bf r}_i$. 
The advantage of the RPC tessellations is that the resulting domains are  guaranteed to be convex polyhedra\cite{Gellatly81}  whose inscribed spheres match the input radii  $\{R_i\}$, see Fig.~\ref{weight_voronoi}(b).

\begin{figure}[t]
\centering
\includegraphics[width=8.25cm]{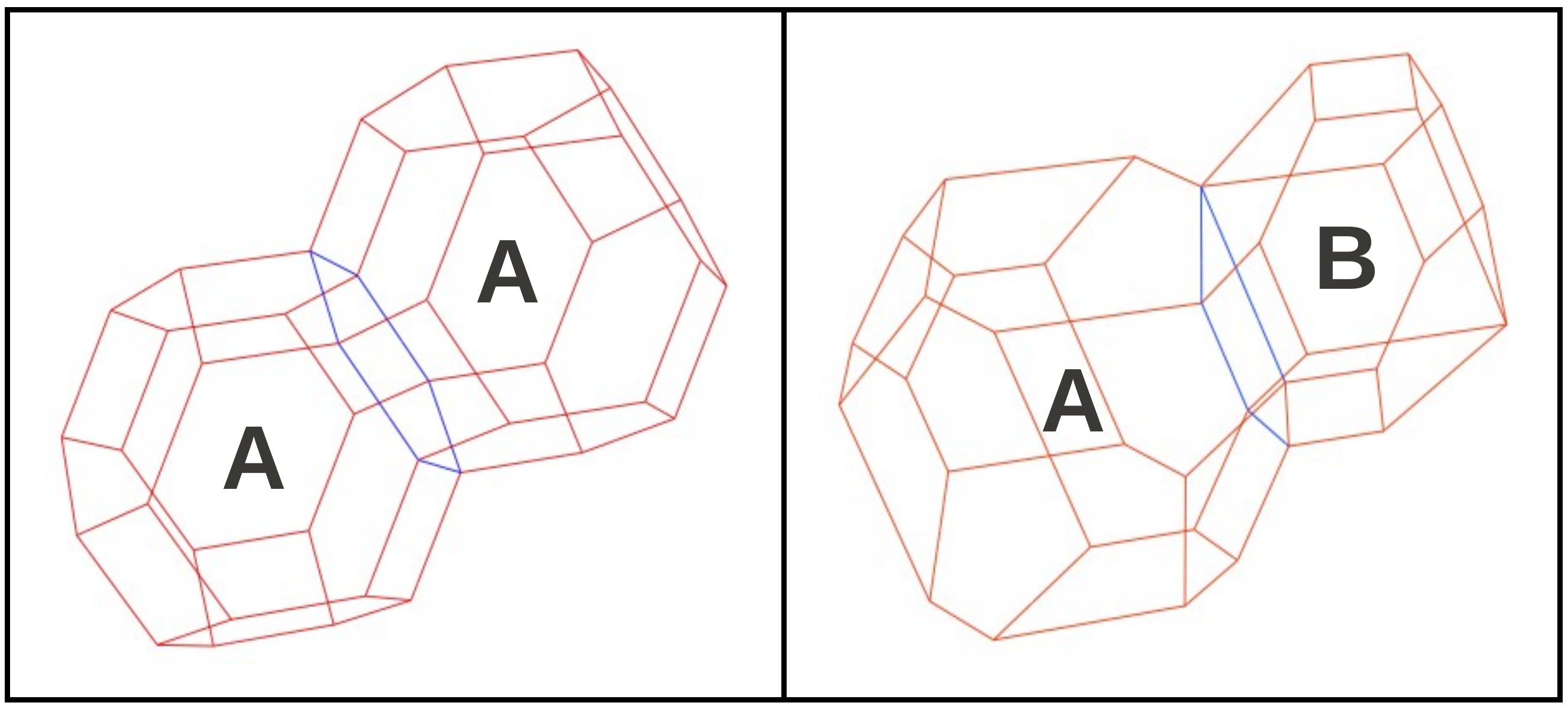}
\caption
{(Color online) VP for 2-atom unit cells with two inscribed radii. (Left)  B2 cell with equal  radii each of type A, and (Right) BCT cell with unequal radii of type A and B.}
\label{polyhed}
\end{figure}

\subsection{Physics-Based Definition of Atomic Size }\label{SPR} 
{\par} To provide the weights  (radii) for RPC, we must choose the ``size'' (radii) of each atom. 
A simple choice is the atomic radii from empirical or theoretical tables,\cite{Slater64-Clement67} which, however, is not site specific nor does it reflect bonding.
For electronic-structure use, a judicious choice for each site (atomic or empty) are the minimum radii selected from the set of saddle-point radii (SPR) in the total electronic charge density, which reflects atomic ``size''.\cite{Alam09}
Initiating a calculation, these SPR can be estimated by overlapping the isolated-atom charge densities in the desired structural positions, similar to L\"owdin potentials.\cite{Alam09} 
For a spherical-harmonic basis, we have shown that this SPR representation, even within atomic-sphere approximation, dramatically improves  basis-set convergence and energetics in size-mismatched systems compared to full-potential methods.\cite{Alam09,Alam10} 
Site-charges  now also obey electronegativity rules, as found also with Bader topological (non-convex) cells.\cite{Alam09,Alam10} 
Full details of applications are available in Ref.~\onlinecite{Alam09}.

{\par} For RPC, given the site locations (structure) and SPR (weights), we use our modified version of Bernal's \emph{Fortran} software\cite{Bernal} to generate the VP information (vertices, faces, edges etc.).
Figure~\ref{polyhed} shows the VP generated for a B2 cell with atoms of equal size and a tetragonally-distorted BCT cell with atoms of unequal size.

\subsection{Dual Coordinate Transformation and Gauss Quadrature Sums}
\begin{figure}[t]
\centering
\includegraphics[width=7.5cm]{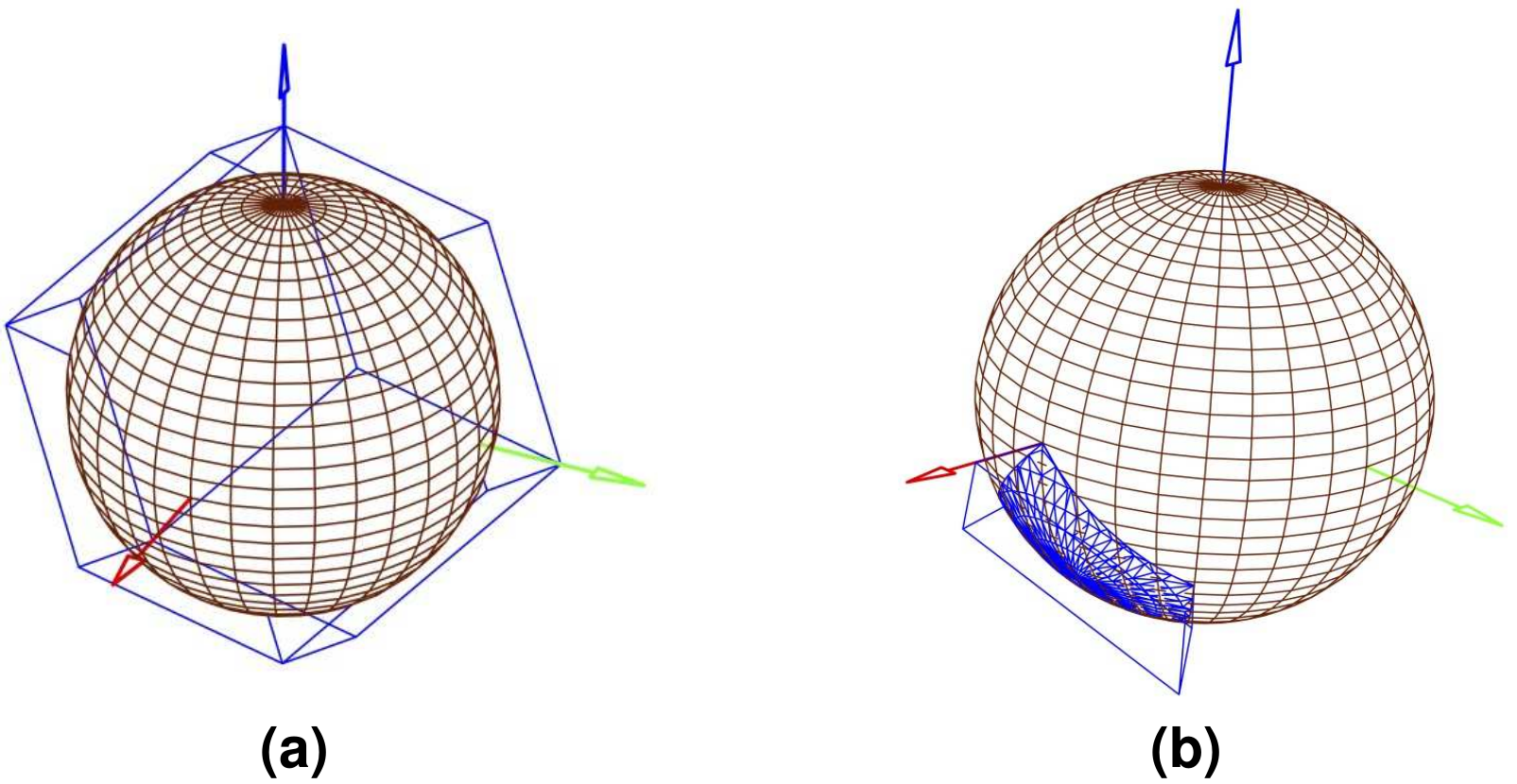}
\caption
{(Color online) (a) VP of a FCC structure with twelve quadrilateral faces and an inscribed (touching) sphere. (b) A section of the VP shown as single truncated pyramid.}
\label{VP_FCC}
\end{figure}

Having divided the system into VP, there are two ways to proceed depending on the nature of the integrand $f({\bf r})$. 
For simple integrands, separate each VP integration over a numerous simple polyhedra associated with each VP face and perform Gauss quadrature sum, and the method works straightforwardly.
If $f({\bf r})$ has singularities near the origin, or if it is accessible only on a sparse grid, then two major VP subdomains need to be handled separately, i.e. inside and outside of the inscribed sphere.
If $f({\bf r})$ is spherical, the integral is one-dimensional and  easy to perform accurately, whereas the second, interstitial domain is more challenging, see Fig.~\ref{VP_FCC} for FCC example. 

{\par} The interstitial has too unusual a boundary for the direct determination of suitable sampling points and their weights. 
 To find the sampling points, we transform a bi-unit cube $-1 \le x,y,z \le 1$ into each pieces of the interstitial formed by each VP face and the site center but cropped by the inscribed sphere.
If any face has more than four vertices, points are added within the face (uniformly distributed) so that each face can be subdivided into polygons always having at most four vertices (a quadrilateral base);  as a result, no interstitial subdomain has more than eight corners, like the cube.
The same map used on the Gauss-Legendre points tells us the sampling positions in each interstitial subdomain. 
Note that one could use a triangular base, but we find that, while both subdivisions give the same results, the quadrilateral requires less operations, hence, it is more efficient. 

%

{\par} For clarity, consider a one-atom FCC crystal, as in Fig.~\ref{VP_FCC}(a), where the VP consists of 12 quadrilateral faces, which are divided into 12 cropped pyramids. 
Pick one, as in Fig.~\ref{VP_FCC}(b), and introduce spherical coordinates ($r,\theta,\phi$) so that the $z$-axis is perpendicular to the VP face. 
Within each piece, the radius $r$ runs from the inscribed radius $R$ to the pyramid base (or VP face). 
To consider the case where the inscribed sphere integral is not done separately, take $R\rightarrow 0$ in what follows, and each pyramidal piece will no longer be cropped.

\begin{figure}[t]
\centering
\includegraphics[width=7cm]{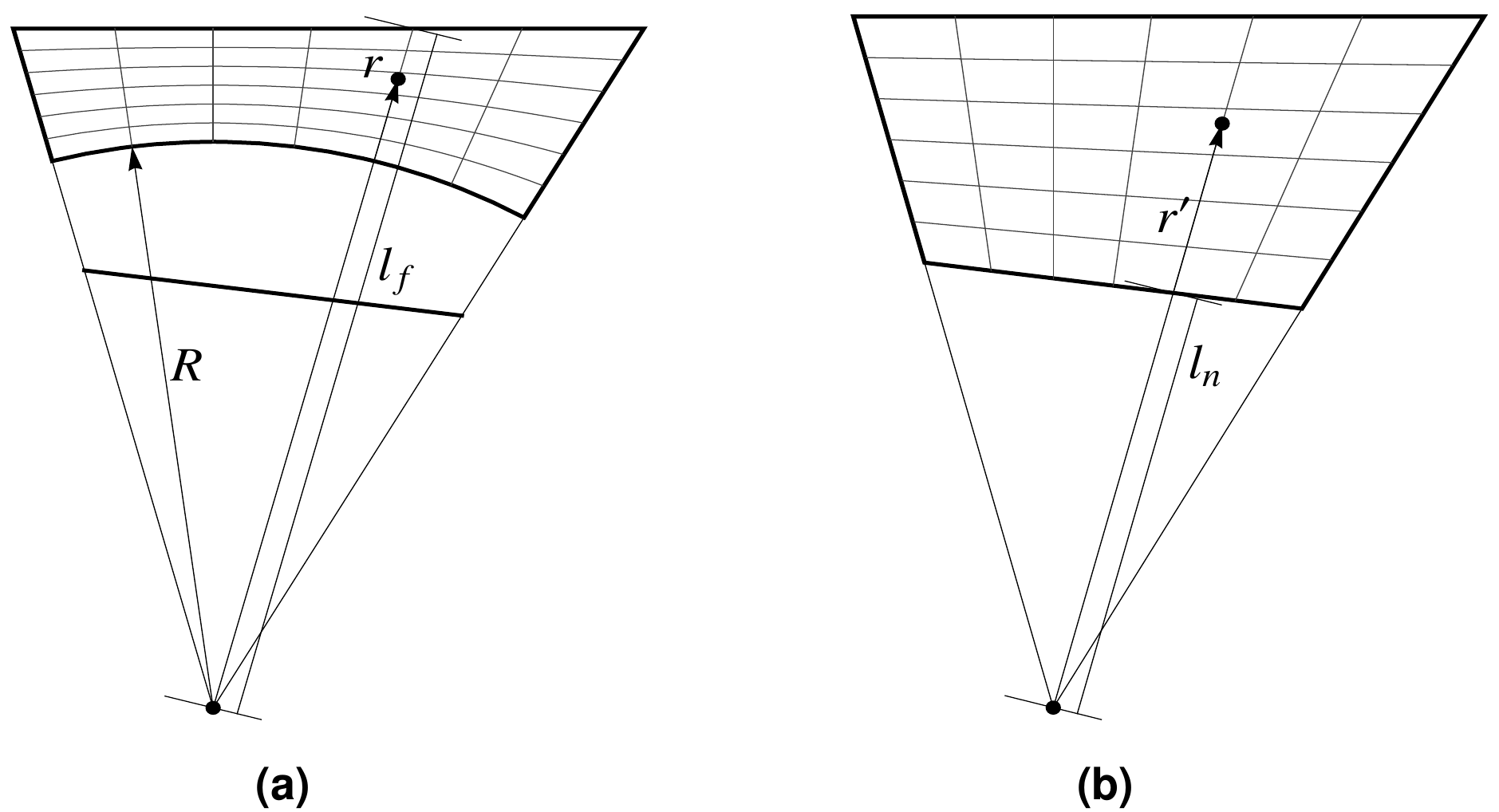}
\caption
{Cross-section of the cropped pyramid (a) before radial scaling and (b) after radial scaling. }
\label{flatten}
\end{figure}

{\par} Before we map the cube to this element, we must find a transformation that flattens the curved interior surface. 
Choose any three of the four corner vertices formed by the intersection of the pyramid and the inscribed sphere. 
 These three points are taken to define an interior plane. 
 Now consider a cross-section of the element at fixed angle $\phi$ or $\theta$, which resembles Fig.~\ref{flatten}(a). 
 Note $l_n$ is the distance from center of the inscribed sphere to point of intersection of radius vector with interior plane; and $l_f$ is the distance  to intersection with base plane (or face). 
 Then the map
\be
r = \frac{1}{l_f-l_n}\left[\rule{0mm}{4mm} l_f(R-l_n) + r' (l_f-R)  \right]
\label{eq3}
\ee
will radially expand the interstitial piece (unprimed coordinates) so that the surface cut of the inscribed sphere will map to the interior plane  (primed coordinates). 
Note that the map as given takes the plane to the sphere, because, ultimately, we want a map from the cube to the interstitial piece. 
 Despite the simplicity of the map (Eq. \ref{eq3}), the Jacobian $J_1$ is non-polynomial due to the angular dependence of $l_f(\theta,\phi)$ and $l_n(\theta,\phi)$. 
 The standard determinant form of $J_1$ can be simplified by considering the volume change of an infinitesimal cell embedded in a spherical  coordinate mesh. 
The cell will be stretched radially by a factor of  $dr/dr' = (l_f-R)/(l_f-l_n)$. 
And, because the cell will be translated radially  from $r'$ to $r$, the base area will change from ${r'}^2 d\Omega$ to $r^2 d\Omega$. 
Thus, the total volume change (ratio) of the cell will be $\frac{(l_f-R)}{(l_f-ln)} \frac{r^2}{{r'}^2}$.

\begin{figure}[t]
\centering
\includegraphics[width=8.25cm]{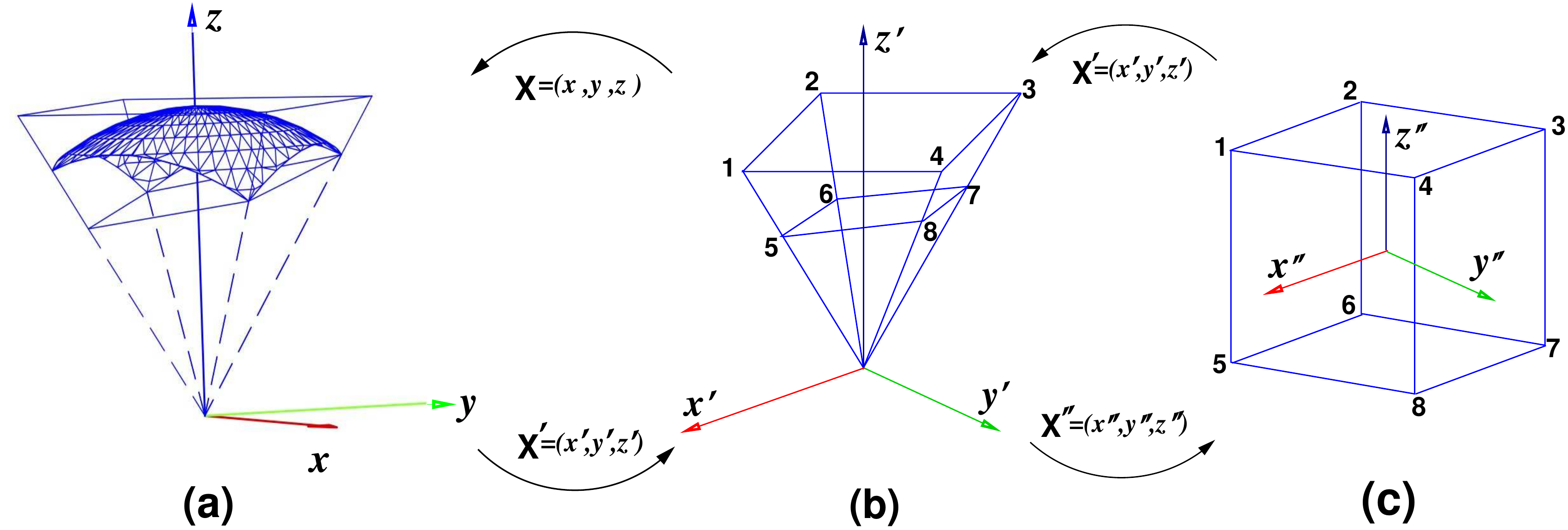}
\caption
{(Color online) Two-step coordinate transformation: (1) Bottom curved surface (a) to the interior plane (b) via the Jacobian $J_1$, and (2) hexahedra in (b) to the isoparametric ($2\times 2\times 2$)  bi-unit cube (c) via the Jacobian $J_2$.}
\label{lateral_hex_biunit}
\end{figure}

{\par} Having flattened the interior, curved surface, we then perform a second mapping from this hexahedra to a bi-unit $2\times2\times2$ cube, as depicted in Fig. \ref{lateral_hex_biunit}. 
Let ($x',y',z'$) and ($x'',y'',z''$) be the  coordinates before and after the transformation, respectively. 
Mathematically, we can connect them using the expression
\begin{widetext}
\be
\left[  \begin{array}{ccc} x' & y' & z'  \end{array} \right]
 = \frac{1}{8} 
\left[  \begin{array}{cccccccc} 
1 & x'' & y'' & z''  & x''y'' & y''z'' & x''z'' & x''y''z''
\end{array} \right]
\left[  \begin{array}{cccccccc}
\ 1 &\ 1 &\ 1 &\ 1 &\ 1 &\ 1 &\ 1 &\ 1 \\
\ 1 & -1 & -1 &\ 1 &\ 1 & -1 & -1 &\ 1 \\
 -1 & -1 &\ 1 &\ 1 & -1 & -1 &\ 1 &\ 1 \\
\ 1 &\ 1 &\ 1 &\ 1 & -1 & -1 & -1 & -1 \\
 -1 &\ 1 & -1 &\ 1 & -1 &\ 1 & -1 &\ 1 \\
 -1 & -1 &\ 1 &\ 1 &\ 1 &\ 1 & -1 & -1 \\
\ 1 & -1 & -1 &\ 1 & -1 &\ 1 &\ 1 & -1 \\
 -1 &\ 1 & -1 &\ 1 &\ 1 & -1 &\ 1 & -1 
\end{array} \right].
\left[  \begin{array}{ccc}
 x_{1}' & y_{1}' & z_{1}'  \\
 x_{2}' & y_{2}' & z_{2}'  \\
 x_{3}' & y_{3}' & z_{3}'  \\
 x_{4}' & y_{4}' & z_{4}'  \\
 x_{5}' & y_{5}' & z_{5}'  \\
 x_{6}' & y_{6}' & z_{6}'  \\
 x_{7}' & y_{7}' & z_{7}'  \\
 x_{8}' & y_{8}' & z_{8}'  \\
\end{array} \right].
\label{eq5}
\ee
\end{widetext}
where the index in the subscript ($1$ to $8$) indicates the vertex number in Fig.~\ref{lateral_hex_biunit}. 
In this map, we have reverted to describe the hexahedral element in cartesian coordinates $(x',y',z')$ rather than the spherical  $(r',\theta',\phi')$.

{\par} The Jacobian of the transformation $J_2$ that turns the hexahedra into a bi-unit cube is
\be
J_2 = \left|  \begin{array}{ccc}
\frac{\partial x'}{\partial x''} & \frac{\partial x'}{\partial y''} & \frac{\partial x'}{\partial z''} \\
\frac{\partial y'}{\partial x''} & \frac{\partial y'}{\partial y''} & \frac{\partial y'}{\partial z''} \\
\frac{\partial z'}{\partial x''} & \frac{\partial z'}{\partial y''} & \frac{\partial z'}{\partial z''} 
\end{array} \right|.
\label{eq6}
\ee

{\par} Thus, the volume integral over the interstitial region transforms to a volume integral over a cube. 
This can be expressed, using Gaussian-Legendre integration, as
\ber
\int_{\Omega^{IS}} f({\bf r}) d^{3}{\bf r} 
   &=& \int_{-1}^{1}\int_{-1}^{1}\int_{-1}^{1} d^3 {\bf r''} f({\bf r''})\ J_1\ J_2\nonumber \\
   &=& \sum_{l=1}^{N_l} \sum_{m=1}^{N_m} \sum_{n=1}^{N_n}
       f(x_{l}^{\prime\prime},y_{m}^{\prime\prime},z_{n}^{\prime\prime})   
       J(x_{l}^{\prime\prime},y_{m}^{\prime\prime},z_{n}^{\prime\prime})      \nonumber \\ 
   && \qquad \qquad \times    \ w_l(x_{l}^{\prime\prime}) \ w_m(y_{m}^{\prime\prime}) \ w_n(z_{n}^{\prime\prime}) 
\label{gaus_quad} 
\eer
where $J=J_1 J_2$, and $N_l$, $N_m$ and $N_n$ are the number of quadrature points along $x''$-,$y''$- and $z''$-axes, respectively. 
The Gauss points $x_i$ and weights $w_i$ are known analytically from the zeroes of the Legendre polynomial, so Eq.~\eqref{gaus_quad} is straightforward to evaluate.
Calculation time is primarily spent in numerically evaluating the analytically-derived Jacobians $J_1$ and $J_2$ for the two successive transformations and the $f(x'',y'',z'')$, hence, quite fast.
This isoparametric approach achieves machine-precision error for VP integrals involving volume, charge-densities and potentials. 
The function $f(x'',y'',z'')$ should be evaluated at the specified  $x_i$ points; if, however, $f$ is only defined on a discrete grid, the function must be interpolated to each $x_i$, in which case interpolation error is the major error that should be ameliorated to achieve high-accuracy integration.
Generally, if $f(x'',y'',z'')$ is a polynomial of order $p_1$, $p_2$ and $p_3$ along the three directions, respectively, then the number of sampling points $N$ required to integrate the quantity exactly for a simple polyhedra domain is $(\frac{p_1}{2}+1)\times(\frac{p_2}{2}+1)\times(\frac{p_3}{2}+1)$.
For the case where we separate the integral over the inscribed sphere and integrate the interstitial over a domain that is curved, the transformation makes the integrand effectively non-polynomial; therefore, more Gauss points will be required.

{\par} Our method is distinguished from that in Refs.~\onlinecite{Boerr88} and \onlinecite{Velde92} by the choice of transformations, as well as partitioning space via the weighted VP. 
We transform the Gauss-Legendre sampling points inside a bi-unit cube into the truncated pyramid by (1) cubic polynomial mapping of the corner points of the cube to the corner points of the truncated pyramid (given by $J_2$), and then (2) performing a linear mapping (in radius) of the interior plane (or side closest to origin) onto the relevant cut of the inscribed sphere (given by $J_1$). 
Our Jacobian $J \equiv J_1 J_2$ is always smooth and well-behaved, even for highly skewed pyramid. 
Baerends {\it et al.}\cite{Boerr88, Velde92} have noted that their choice of coordinates can cause their intermediate functions to behave poorly (i.e., the $J$ diverges) when the pyramid has wide opening angles, or a strongly skewed face.
For our $J$ to diverge, the interior plane would need to (nearly) touch the pyramid base plane; but, with the interior plane defined as the one passing through three of the intersection points of the inscribed sphere and the edges of the pyramid, this could only happen if the sphere touched one of the corners of the VP, which can never happen. 
In addition, the present procedure requires minimally fewer function evaluations.

\subsection{Symmetry Considerations} 
{\par} We could take advantage of the symmetry of the VP and crystal. 
We consider two kinds of symmetry. 
First, the point-group symmetry of the crystal structure identifies the set of inequivalent sites in the cell, which reduces computational time to that over inequivalent sites only.
The second symmetry is  associated with individual polyhedra. 
Because each polyhedra consists of various quadrilaterals  associated with their faces, we can identify the symmetry equivalent pyramids by applying a set of symmetry operators over each polyhedra around the symmetry unique atoms. 
By integrating only over symmetry-inequivalent pyramids corresponding to each symmetry-inequivalent atoms and weighting them with their degeneracy, an appreciable savings in computer time would be obtained for systems with high symmetry.
For example, there are $12$ facets for VP in an elemental FCC structure, so, at a minimum, we could perform an integration over one VP facet and multiply result by $12$; however, because each facet is $4$-fold symmetric, we could do $1/4$ of the VP facet and multiply result by $48$, reducing VP integration by $1/48$ of above integrations timings.

\subsection{All-Electron Implementation}
Our  method is conceptually simple and easy to implement in any general electronic-structure code, with additional advantages for site-centered methods.
For a spherical-harmonic basis, integrating separately over the inscribed sphere directly eliminates the Coulomb singularities due to the Jacobian within spherical coordinates.
Also, using the  optimal SPR basis\cite{Alam09} we have better basis-set convergences and site charges.
We have included this isoparametric method in a small set of RPC routines to determine the VP (vertices, faces, and edges).
For complex cells, our algorithm has the flexibility to control the desired precision to balance the computational cost. 

{\par} This software is used to implement the (un)weighted VP-based isoparametric integration in our all-electron, KKR-CPA  code.\cite{MECCA} 
We discuss these results in  Section \ref{results}.
Details of the calculations are as follows.
The Green's function and related integrals use an external angular-momentum cutoff up to L$_{max}=3$, i.e., $s-$, $p-$, $d-$, and $f-$symmetries, as needed.
Energy integrations of the Green's functions are done by contour integration\cite{Johnson-fastKKR} via Gauss-Chebyshev methods with $18$ energy points.
We use the local spin-density approximation (LSDA) as parametrized by von-Barth-Hedin.\cite{vonBH}
The Brillouin zone integrations use the Monkhorst and Pack\cite{Monkhorst} special k-point method with $20^3$ points in the full zone. 
For disordered alloys, we use the coherent-potential approximation\cite{KKR-CPA-Johnson} (CPA) based on the screened-CPA\cite{scr-cpa} to incorporate more properly the metallic screening due to charge correlations in the local chemical environment.

\section{An Exactly Solvable Model} \label{comput_det}
To illustrate the numerical convergence and accuracy, we use Van-Morgan's exactly solvable charge-density model.\cite{Morgan77} 
Many standard electronic-structure kernels can be exactly evaluated for the van Morgan density and potential, so the error in the numerical integrals can be precisely determined. 
We verify that accurate results are found with a modest number of Gauss points that depend on structure, and machine-precision can be achieved by increased number of  points,  slightly increasing computational time.
 
{\par}We showcase the convergence of \emph{volume} and \emph{charge} conservation, the [$\rho({\bf r}) V({\bf r})$] integral evaluated for \emph{kinetic and/or Coulomb} energy, and more highly varying functions in $l$ and $\mathbf{r}$.
Apart from the cubic structures, we have also tested the convergence of the interstitial volume integral for more complex crystal structures.
In the timings below, we have not utilized the associated symmetry of the crystal and the VP, so that the results reflect the most inequivalent case.

\subsection{The van Morgan Test Problem} 
{\par}  The van Morgan \cite{Morgan77} test charge density is defined as
\be
\rho({\bf r}) = B\sum_{n=1}^{K} e^{i\ {\bf T_{n}.r }},
\ee
where ${\bf T}_{n}$ are the nearest-neighbor reciprocal lattice vectors, and $B$ is a scale factor.
We will take $B=1$ for simplicity. 
(From the Bauer expansion, a plane wave requires, in principle, an infinite number of spherical harmonics to be fully represented.) 
Because $\Omega^{VP}$ and $\Omega^{MT}$ are known exactly for any crystal structure, it is often convenient, especially for site-centered methods, to divide the VP into two volumetric regions: the volume of inscribed sphere $\Omega^{MT}$ and the volume within the interstitial region $\Omega^{IS}$, so that $\Omega^{VP}= \Omega^{IS}\cup \Omega^{MT}$.

{\par} First, we can precisely assess the numerical error associated with \emph{volume conservation} via
\be
\int_{IS} d^{3}r = \Omega^{IS}=\Omega^{VP}-\Omega^{MT},   
\ee
where $\Omega^{MT}=4\pi R^3/3$, and, for example, $R$ is $1/2$, $\sqrt 2/4$, and $\sqrt 3/4$ for SC, FCC, and BCC (in units of lattice constant), respectively.
The left-hand-side numerical integral is compared with the analytical result available for the right-hand side.
For example, the VP volumes are $1$, $1/4$, and $1/2$ (in units of lattice constant cubed) for SC, FCC, and BCC, respectively. 

{\par}Second, we can assess the integrations associated with \emph{charge conservation}, including the determination of electronic chemical potential or Fermi energy.
With $\rho({\bf r})$ having no zero-mode component in its Fourier expansion, the integral of charge over a VP cell must be identically zero; hence, charge neutrality requires that
\be
Q^{\text{total}}=\int_{VP}\rho({\bf r}) d^{3}r = 0.   
\ee
Subdivision of VP yields
\be
Q^{IS}=\int_{\Omega^{IS}}\rho({\bf r}) d^{3}r = -   \int_{\Omega^{MT}}\rho({\bf r}) d^{3}r.
\ee

{\par}Next, we can assess numerical errors for the $\rho({\bf r})V({\bf r})$ integral, which can be expressed as
\begin{eqnarray}\label{eq:rhoV}
[\rho V]^{IS}&=&\int_{\Omega^{IS}}\rho({\bf r}) V({\bf r}) d^{3}r\nonumber\\
 &=& \frac{4 \pi K \Omega^{IS}}{|T_n|^2} - \int_{\Omega^{MT}}\rho({\bf r}) V({\bf r}) d^{3}r.
\end{eqnarray}

{\par} The exact analytic solutions of the above kernels for three cubic structures (SC, FCC and BCC) are given in the Appendix \ref{app1}. 
However, non-cubic structures are numerically no more difficult or error prone than these cubic cases (but they cannot be performed analytically).
Besides band-energy (an eigenvalue summation requiring a Fermi energy) and exchange-correlation, the above three integrals reflect the main integrations contributing to DFT total energies, for example.

\subsection{Complex Varying Integrands}\label{complic_func}
The general method we have presented here can integrate over any arbitrary polyhedra for any complicated function, such as those with high angular momentum or strongly varying with(out) exponential decay.
Here we showcase a set of strongly varying integrands that are critical for evaluating the near-field contributions to the Poisson equation (in the so-called \emph{moon region}) for site-centered, electronic-structure methods,\cite{Gonis-Butler} i.e.,
\begin{equation}
a_{lm}=\sum_{R\ne 0}\int_{VP} d{\bf r'} \rho_{\bf R} ({\bf r'}) \frac{Y_{lm}(\widehat{\bf r'+R})}{\vert {\bf r'+R}\vert^{l+1}}
\label{a_L}
\end{equation}
where $Y_{lm}$ are the spherical harmonics and there is a Madelung summation over {\bf R}. 
It is a rapidly decaying function with increasing $l$, and, to achieve high precision of this  piece of the Coulomb potential, rather high $l$'s are required. 
Convergence of the above integral for various $\{lm\}$-values is shown in the next section. 
We have also tried other more strongly varying functions, and again achieved accurate results with modest number of Gauss points.
As will be discussed elsewhere, most codes that implement the correct Poisson solution for space-filling VPs cannot do the integrals for skewed VP, or they are not accurate enough due to use of, e.g., shape-functions, an example appears below.


\begin{figure}[t]
\centering
\includegraphics[width=8cm]{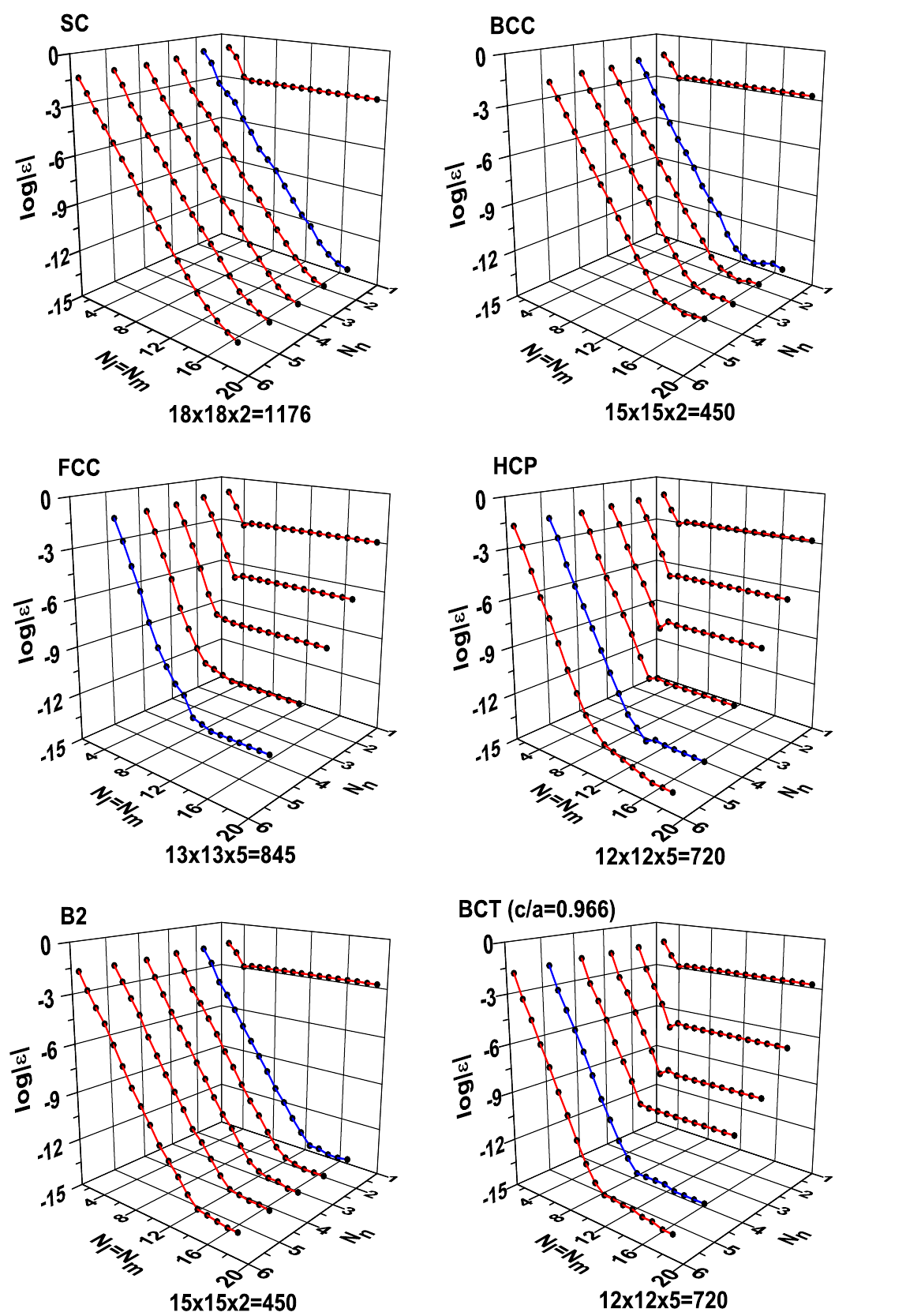}
\caption
{(Color online) Logarithmic (base-10) error in interstitial volumes for six structures. $N_i~(i=l,m,n)$ is the number of Gauss points along $\hat{x}''$, $\hat{y}''$ and $\hat{z}''$, respectively, with $N_n < N_l=N_m$ due to a smaller caliper along $\hat{z}''$. Dark (blue) lines indicate minimum number of points along $\hat{z}''$ (total points listed below plots) to achieve $13$~decimal accuracy.}
\label{vol_int}
\end{figure}

\begin{figure}[t]
\centering
\includegraphics[width=8cm]{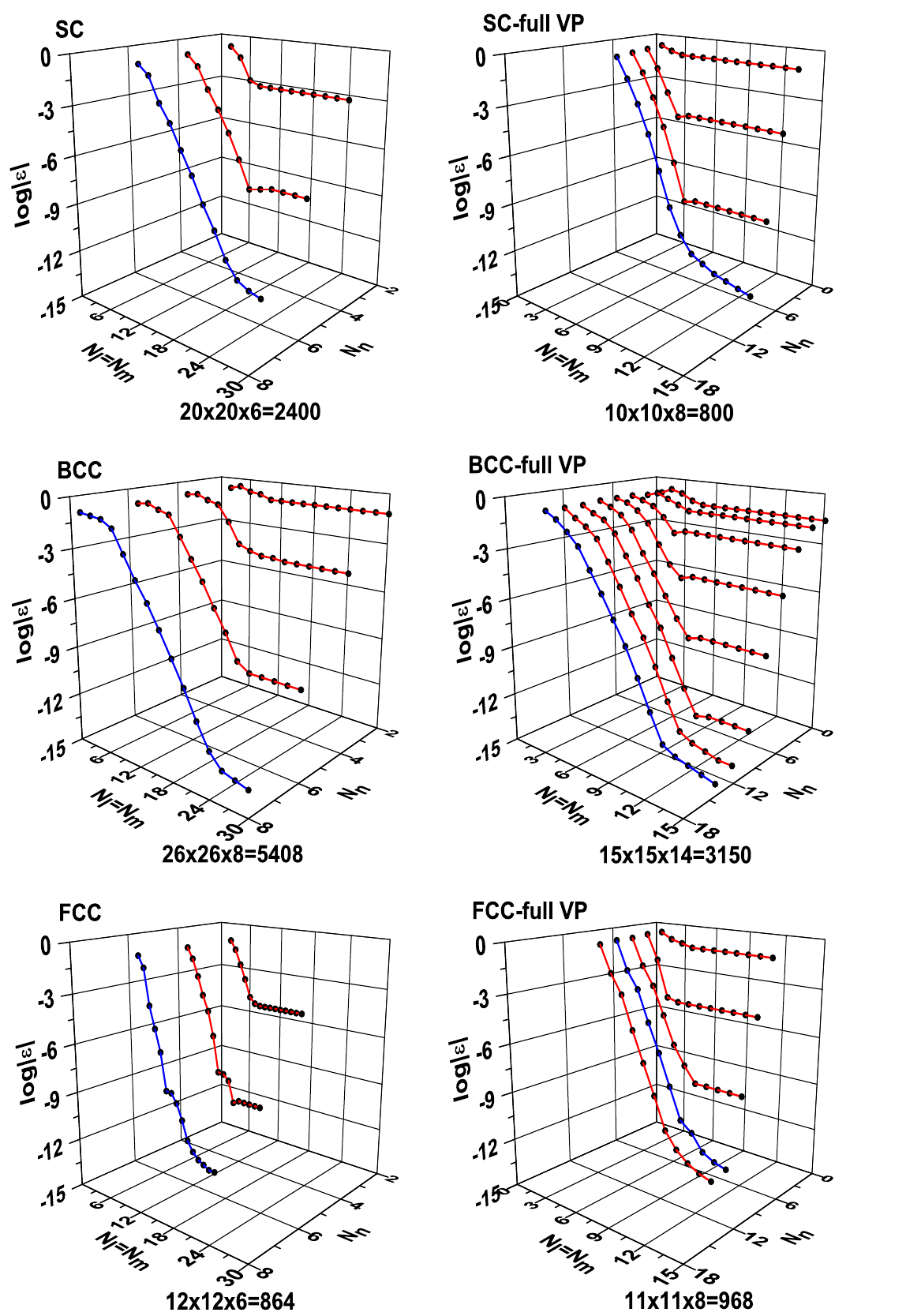}
\caption
{(Color online) For the van-Morgan problem for SC, BCC, and FCC, (left) the logarithmic (base-10) error in the interstitial charge, i.e., $\epsilon=(Q^{IS}_{calc} - Q^{IS}_{exact})/Q^{IS}_{exact}$, and  
(right) absolute error in VP total charge, i.e., $\epsilon=Q^{VP}_{calc} - Q^{VP}_{exact}$. 
Other details are as in Fig.~\ref{vol_int}. }
\label{chg_convg}
\end{figure}

\begin{figure}[t]
\centering
\includegraphics[width=4.5cm]{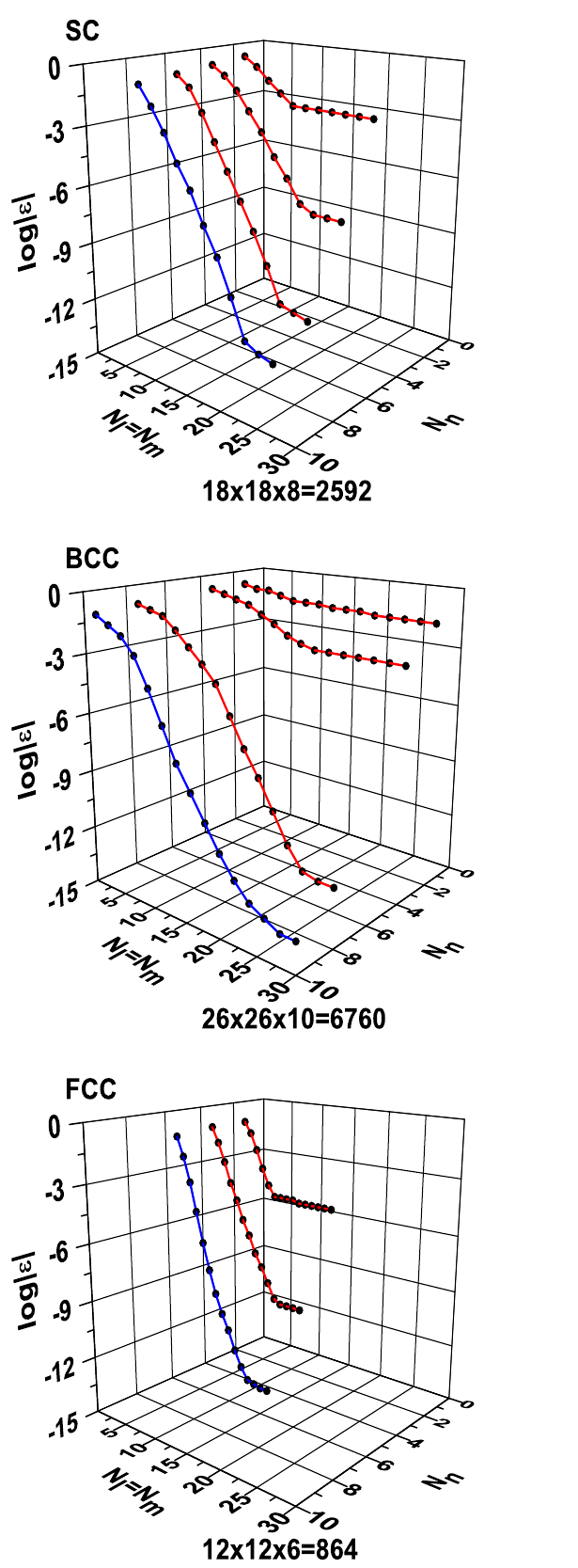}
\caption
{(Color online) For the van-Morgan problem for SC, BCC, and FCC, the logarithmic (base-10) error in the interstitial--[$\rho V$] integral. 
Other details are as in Fig.~\ref{vol_int}. }
\label{rhoV_convg}
\end{figure}

%
\section{Results and Discussion}\label{results}

\subsection{Accuracy} 
{\par} To illustrate the convergence of isoparametric integration,  Figure~\ref{vol_int} shows the logarithmic error in interstitial volume for six structures (i.e., 1-atom cubics, 2-atom hcp, and 2-atom B2 and BCT). 
Each point on the graph represents the result for a combination of quadrature points $(N_l,N_m,N_n)$. 
From Fig.~\ref{lateral_hex_biunit}(a), it is clear that the cropped pyramid has a thinner dimension along the $z$-axis compared to the other two axes. 
Therefore, we use less quadrature points along $\hat{z}''$ than the $\hat{x}''$ and $\hat{y}''$, i.e., $N_n < (N_l,N_m)$; in particular, we used $N_l =N_m$. 
Accuracy of around $10^{-3}$ is already reached with only $N_l=N_m=4 $ points along the $\hat{x}''$ and $\hat{y}''$. 
The darker line in each panel shows the minimum number of quadrature points along $\hat{z}''$ to achieve a convergence to $13$ decimal places. 
For example, the minimum number of Gauss points along $\hat{z}''$ for a BCC structure to attain an error less than $10^{-13}$ is two. 
The minimum number of points $(N_l,N_m,N_n)$ required  is listed below each subpanel. 

\begin{table}[b]
\caption{\label{convg_table1} Convergence for the interstitial volume, charge 
and $[\rho V]$ integrals for various crystal structures. $\{N_l=N_m, N_n \}$ are
 the optimal number of points for each structure to reach an accuracy of at 
least 13-decimal places. $VC$, $QC$ and $[\rho V]$ stands for the volume, 
charge and $[\rho V]$-integral convergence.}
\begin{ruledtabular}
\begin{tabular}{lllll}
\vspace{0.2cm}
Structure & $\{N_l,N_n\}_{VC}$ &  $\{N_l,N_n\}_{QC}$ &  $\{N_l,N_n\}_{\rho V}$ \\
SC   & $ \{18,2 \}$  &  $ \{20,6 \}$  &  $ \{18,8 \} $\\
BCC  & $ \{15,2 \}$  &  $ \{26,8 \}$  &  $ \{26,10 \}$\\
FCC  & $ \{13,5 \}$  &  $ \{12,6 \}$  &  $ \{12,6 \} $\\
HCP  & $ \{12,5 \}$  &                &               \\
B2   & $ \{15,2 \}$  &                &               \\
BCT  & $ \{12,5 \}$  &                &               \\
\end{tabular}
\end{ruledtabular}
\end{table}

{\par} The convergence of the charge density integral ($Q$) is given in Fig. \ref{chg_convg}. 
The left panel shows the logarithmic error in the interstitial charge  $Q^{IS}$ for the cubic structures. 
The right panel shows the absolute error $\epsilon^{VP} = Q_{calc}^{VP}-Q_{exact}^{VP}$ in the total charge integral. 
The charge convergence requires more points to yield a similar level of accuracy. 
For example, to achieve an accuracy of up to the third-decimal place, the BCC structure requires 8-points along the $\hat{x}''$ and $\hat{y}''$ compared to the 4-points needed for the SC and  FCC structures. 
Higher accuracy requires more points for BCC case due to its wider and more asymmetric interstitial region. 

{\par}Figure \ref{rhoV_convg} shows the convergence of the interstitial $[\rho V]$-integral for the cubic structures. 
It is interesting to notice that $[\rho V]$-integral converges almost at the same rate as the 
$\rho$-integral (left panel of Fig. \ref{chg_convg}) and does not take any longer for all the three structures. 
BCC structure in this case also requires comparatively more points to converge for the same reason in the previous paragraph. 
We have analyzed other relevant integrands (often used in electronic-structure calculation) as well
and found either a similar or a slightly slower rate of convergence.

{\par} In Table \ref{convg_table1}, we have listed the minimum number of points required to get the interstitial volume, charge and $[\rho V]$-integral convergence to more than $13^{th}$ decimal for each structure. The number of points required are given as $\{N_l=N_m,N_n\}$.
Table \ref{convg_table2} shows similar results for the full VP integral of strongly varying functions in Eqn.~\eqref{a_L} for various $\{l,m\}$ values, exhibiting oscillatory angular dependence with $l$-dependent spatial decay. 
In spite of its strongly varying nature, the number of Gauss points $\{N\}$ required to achieve an accuracy of up to 10-decimal place is not large, and are comparable to those of $\rho$ and $\rho V$ integrals shown in Table \ref{convg_table1}.

\begin{table}[t]
\caption{\label{convg_table2}  Convergence of fcc $a_{lm}$ in Eqn.~\eqref{a_L} versus $\{l,m\}$ ($R$ is summed to 8$_{th}$ neighbor shell). $\{N\}$ is the number of points per direction for $10$-decimal place accuracy. }
\begin{ruledtabular}
\begin{tabular}{lllll}
\vspace{0.2cm}
$l$  &  $m$  &  $\{N\}_{a_L}$ & $[a_{lm}]_{\text{numerical}}$ &  $[a_{lm}]_{\text{exact}}$ \\ 
\vspace{0.15cm}
0  &  0  &   12 &  ~0.009951109455    &  ~0.009951109341     \\
\vspace{0.15cm}     
2  &  0  &   12 &  ~0.000000000000    &  ~0.000000000000      \\
4  &  0  &   14 & -9.449717387589     & -9.449717387292     \\
\vspace{0.15cm}
4  &  4  &   14 & -5.647286285886     & -5.647286285399     \\
6  &  0  &   16 & -10.62648231314     & -10.62648231381     \\
\vspace{0.15cm}
6  &  4  &   16 &  ~19.88032802119    &  ~19.88032802174      \\
8  &  0  &   18 &  ~65.84024514425    &  ~65.84024514410     \\
8  &  4  &   18 &  ~24.75927136401    &  ~24.75927136434     \\
\vspace{0.15cm}
8  &  8  &   19 &  ~37.72380770670    &  ~37.72380770699     \\
10 &  0  &   21 &  ~135.8520785234    &  ~135.8520785239     \\
10 &  4  &   21 & -136.8931058432     & -136.8931058438     \\
\vspace{0.15cm}
10 &  8  &   24 & -162.9353862901     & -162.9353862928     \\
12 &  0  &   24 & -205.2050307885     & -205.2050307878    \\
12 &  4  &   26 & -493.6493920838     & -493.6493920829     \\
12 &  8  &   26 &  ~544.8113627129    &  ~544.8113627135     \\
12 & 12  &   26 & -261.8046415883     & -261.8046415883     \\
\end{tabular}
\end{ruledtabular}
\end{table}

{\par} The accuracy of all our integrals is limited by the accuracy of the VP boundary (vertices, faces and edges) information generated from the Bernal's software.\cite{Bernal} 
We have modified Bernal's original (binary-math/single-precision) code  to improve its efficiency and extend its accuracy, and we were able to achieve just below $10^{-13}$. 
We have verified that our main limitation in accuracy is due to lack of a \emph{double-precision real} code.
By rewriting the software from scratch, which is a considerable effort beyond the scope of present work, we could certainly achieve machine precision. 
Therefore, all integration results will be limited to just below $10^{-13}$; with improved accuracy of VP information, machine-precision is achievable with similar Gauss points described.  

\subsection{Efficiency} 
{\par} To contrast the VP construction timings, we compare to the time required to expand the shape function (or 3-D step function) into spherical harmonics.\cite{wang94} 
The shape-function approach is often used in the community when needing site-dependent quantities. 
The EMTO, KKR, LSMS, APW, etc., methods, for example, typically reports site-quantities, and KKR Green's function methods require site-dependent VP scattering matrices.

{\par}The shape-truncated function for a VP is defined as
\be
 \sigma({\bf r}) = \left\{ \begin{array}{ll}
1 &  {\bf r\in\Omega }\\
0 &  {\bf r\not\in\Omega }\end{array} \right. 
\label{shape}
\ee
where $\Omega$ is the VP region. The expansion of $\sigma({\bf r})$ in
 spherical harmonics yields the angular momentum decomposition
\be
\sigma_{L}(|{\bf r}|) = \int_{\widehat{{\mathbf r}}} d\widehat{{\mathbf r}} \  
Y_{L}^{*} (\widehat{r}) \ \sigma ({\bf r}) \equiv \sigma_{L}(r) , 
\label{shape_L}
\ee
where the integration is over the angles $\widehat{{\mathbf r}}\equiv$ ($ \theta, \phi$) and $L\equiv (l,m)$. 
The shape function is used to simplify the numerical integration of any function $f({\bf r})$ over the polyhedron volume $\Omega$ as
\begin{eqnarray}
F &=&\int_{\Omega} f({\bf r}) \sigma({\bf r})d^3 r  \nonumber \\
  &=&\sum_{L=0}^{L_{max}} \int dr\ r^2 
\sigma_{L}(r)\int_{\Omega} d\Omega~Y_{L}(\widehat{r}) f({\bf r}) ,
\end{eqnarray}
especially if it is well-represented by spherical harmonics. 

{\par}The expansion coefficients $\sigma_{L}(r)$ must be truncated at a very high $L_{trunc}>>L_{max}$ to achieve an accurate representation of the VP shape and to obtain a reliable integral value.
For example, for FCC structure,  $\rho({\bf r})$ is well represented using $L \le 8$ (i.e., $L_{max}=8$), but the shape-function should have $L_{trunc}>>4 L_{max}$ to have converged $\sigma_{L\le8}(r)$ that will yield an accurate integral. 
As we shall see, this $L_{trunc}$ will limit the accuracy of the integrals in the codes that use this approach, making the shape-function approach unacceptable for general (non-high-symmetry) structures, where $L_{trunc}$ should be significantly larger than in the cubic cases to achieve the same level of accuracy as FCC.

\begin{figure}[t]
\centering
\includegraphics[width=8.5cm]{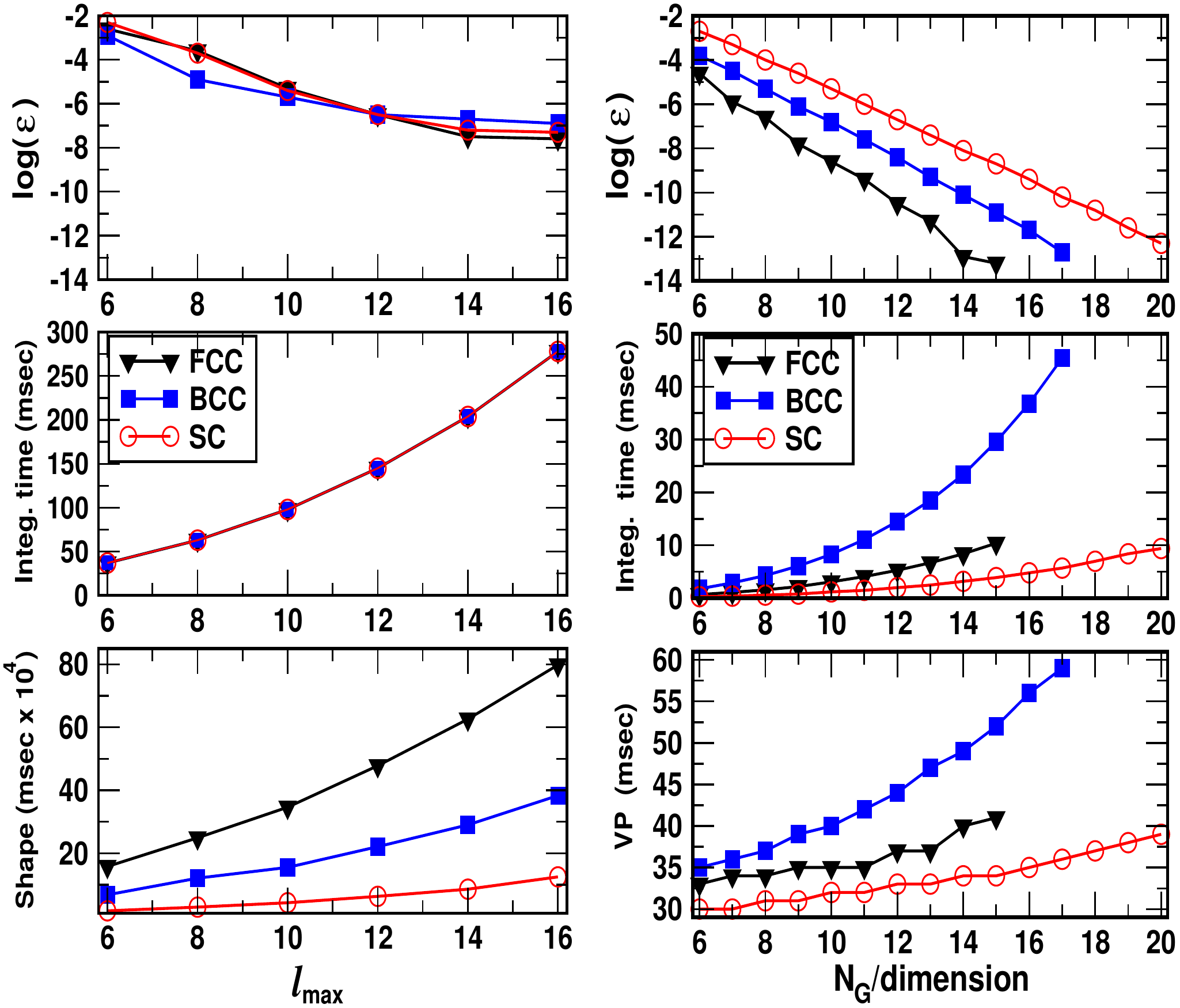}
\caption
{(Color online) Timings to achieve a specific level of interstitial-charge accuracy for cubic structures using shape-function (left) versus isoparametric (right) integration.
Shown in panels are logarithmic error in the interstitial charge (top), and times to construct VP boundary information (bottom) and to integrate (middle). Isoparametric integration is $>10^{5}$ faster and achieves machine precision.}
\label{compare}
\end{figure}

{\par}Figure~\ref{compare} shows accuracy and computer time for isoparametric (right panel) and shape-function (left panel) methods for SC, FCC, and BCC, for a direct comparison.
 The rate of convergence is given with respect to the number of Gauss points along each dimension for the present method, and  with respect to the $l_{max}$ for a fixed radial grid using shape-functions.  
The present method attains error in the van-Morgan interstitial charge below $10^{-13}$ with less computational time.
The shape-function technique cannot achieve an accuracy better than $10^{-7}$ with $l_{max}=16$, an extremely expensive calculation due to the high-$L$ expansion.
Hence, our method provides some significant advantages over existing approaches.

{\par} The bottom panel shows the time required to generate the boundary informations necessary to achieve a certain level of accuracy. 
For both methods, most of the time is spent in determining the VP boundaries. 
The present method generates this information in terms of neighbors, vertices, faces and edges for each VP. 
The shape-function method gets the VP shape in terms of an $L$-expansion on a specific radial grid.
Clearly, the shape-function method requires $>10^4$ more time than the present method. 
The middle panel shows the time (in $m$sec) required to sum the final expression for the integration for both VP or shape function. 
 The present method is faster by $>7$ times.  
Overall, using no symmetry (degeneracy) information to reduce the computational time, we achieve $\simeq 10^5$ faster integration with  $10^6$ less error.

\begin{table}
\caption{Formation enthalpy $\Delta E_f$ (in $m$eV/atom) for (dis)ordered ferromagnetic FePd versus L$_{max}$ from equal (unweighted) and SPR-weighted VPs, along with ordering energies $\Delta E^{o-d}$. SPR results are much less sensitivity to  basis-set L$_{max}$ cutoff. KKR results are compared to other results.}
\label{formation}
\begin{ruledtabular}
\begin{tabular}{lcccccccl}
 L$_{max}$ &   \multicolumn{3}{c}{Unweighted VP}  &  \multicolumn{3}{c}{Weighted VP}   \\
                   & ${\Delta E}_{f}^{ord}$  & ${\Delta E}_{f}^{dis}$ & ${\Delta E}^{o-d}$  
                   & ${\Delta E}_{f}^{ord}$  & ${\Delta E}_{f}^{dis}$ & ${\Delta E}^{o-d}$  \\
\hline
  2     &  $-10.7$  & $+18.6$ & $-29.3$ & $-83.3$ & $-59.4$ & $-23.9$ \\
  3     &  $-45.7$  & $+~8.3$ & $-54.0$ & $-88.8$ & $-63.9$ & $-24.9$ \\
  4     &  $-30.7$  & $-~9.9$ & $-20.8$ & $-86.5$ & $-61.7$ & $-24.8$ \\
  \multicolumn{4}{c} {{Expt. (Ref. \onlinecite{Hultgren})}}  &    $-98\pm11$              &   &  & \\    
  \multicolumn{4}{c} {{VASP-LDA(GGA)}}                         &     $+40$($-130$) &   &  &      
\end{tabular}
\end{ruledtabular}
\end{table}

\subsection{Application to FePd} 
{\par}  We present the formation enthalpy, $\Delta E_f$, for FCC-based ferromagnetic (FM) (dis)ordered Fe-50\%Pd, i.e.,  A$1$ solid-solution and ordered L$1_0$, from an unweighted (equal) VP and SPR-weighted VP, giving an optimal partitioning of space for integration and concomitantly improved basis set, as described in Sec. \ref{SPR} and shown in Fig. \ref{polyhed}.   
Formation energies are highly relevant for phase stability; see Ref.~\onlinecite{Alam10} for  an example application to  phase stability of magnetic-storage materials.

{\par} Table \ref{formation} shows the $\Delta E_f$ for FM L$1_0$ and A1 phases versus the external L$_{max}$ for the local spherical basis.
Using the weighted-VP integration, results become significantly less sensitivity to L$_{max}$, especially clear for the energy difference $\Delta E^{o-d}$ between ordered and disordered phases, which remain almost constant, in contrast to the unweighted case.
Our weighted-integration yields formation energetics in very good agreement with that observed for L$1_0$.\cite{Hultgren} 
The weighted-VP integration thus provides accurate results, a minimal basis set in terms of angular 
momentum cutoff, and a significant reduction in matrix-inversion time because of the now-permitted use of the lower rank of the KKR matrices N(L$_{max} + 1$)$^2$, where N is the number of atoms in the unit cell.

{\par}We could not find enthalpy data for the FM-A1 phase.
Therefore, we provide $\Delta E^{o-d}_{PM}$ in the paramagnetic phase (via disordered local moment state), which is related to order-disorder temperature.\cite{Alam10}
Indeed, for the weighted-VP case, $\Delta E^{o-d}_{PM}$ is $82~m$eV (or $952~$K), close to the order-disorder temperature of 1050~K,\cite{Wang2004} and showing that the disordered phase results are now accurate, too.

{\par} Large-scale boundaries are critical in material science, e.g., for mechanical properties as pinning centers for mageto-anisotropy. 
As a test of the present weighted-VP integration, we calculated the [001] anti-phase boundary planar defect energy of L$1_0$-FePd, a long-period structure with $32$ atoms per cell with varying interstitial regions. 
We find $910~m$J/m$^2$ versus $890~m$J/m$^2$ from VASP plane-wave calculations, only ours is about $20$ times faster, and provided local properties directly.

{\par}In addition, the magnitude of site excess (or deficient) charge, i.e. the ``charge transfer'', in a solid crucially depends on the way in which the space is divided into geometric  cells. 
For space-filling VP, SPR-weighted cells will provide a more  physics-based partitioning of space and more realistic assessment of charge transfer.
Approximate methods like the popular atomic-sphere approximation (ASA) has an overlap error; the situation becomes worse for non-close-packed materials.

{\par} In Table \ref{excess_charge}, we show the calculated  excess charges within (un)weighted VP  sites in A$1$ and L$1_0$ FePd, with comparison to (un)weighted ASA spheres used in many popular codes.
Reference \onlinecite{Alam09} provides details of  ASA approach.
Generally, there is a charge transfer from Fe to Pd, as expected from the electronegativities. 

{\par} However, for unweighted cases in the A1 phase, there is an excess charge on small (Fe) atom, distinctly unphysical, and due to the tails of the charge density of large (Pd) atoms being improperly cut off at the smaller radii.
When a weighted VP or ASA is used, this situation is corrected (the sign changes) because the charge density is now better represented in the disordered phase.\cite{Alam09}
For the unweighted L$1_0$ case, there is a large depletion of charge on small (Fe) atom due to a Madelung effect; however, for the weighted case, the inscribed sphere reflects more appropriately the extent of the charge density and, hence, it is a more reliable estimate.
Importantly, there is a large reduction in excess (depleted) charges for the weighted-VP integration compared to the weighted-ASA case (now with the correct sign), which shows the error arising from overlap of spheres.  

\begin{table}
\caption{Excess charges within VP or ASA spheres for (dis)ordered FePd with (un)weighted  VP via KKR.}
\label{excess_charge}
\begin{ruledtabular}
\begin{tabular}{lcccl}
Method &   \multicolumn{2}{c}{Unweighted}  &  \multicolumn{2}{c}{Weighted}  \\
             & ${\Delta Q}^{ord}$  & ${\Delta Q}^{dis}$ & ${\Delta Q}^{ord}$  &  ${\Delta Q}^{dis}$  \\
\hline
{~VP~~~Fe}  &  $-0.111$ & $+0.059$ &  $-0.032$ & $-0.026$  \\
{~~~~~~~~Pd}  &  $+0.111$ & $-0.059$ &  $+0.032$ & $+0.026$  \\
{ASA~~Fe}   &  $-0.139$ & $+0.089$ &  $-0.082$ & $-0.051$  \\
{~~~~~~~~Pd}                           &  $+0.139$ & $-0.089$ &  $+0.082$ & $+0.051$  \\
\end{tabular}
\end{ruledtabular}
\end{table}

\section{Summary}\label{conclusion}
{\par} We have presented a fast, accurate, and easy to implement method for the numerical integration over general VP for polyatomic systems. 
The algorithm combines a weighted Voronoi partitioning of space with isoparametric integration using the Gauss-Legendre quadrature formulas of product type, and does not suffer from any ill behavior with shape of VP.
In contrast to  other methods, accuracy and convergence was tested rigorously via an analytic charge-density model, with  machine-precision accuracy for reasonable number of Gauss points.
We showed also that our algorithm is $10^5$ faster and $10^{7}$ more accurate than that based on shape-functions used in several electronic-structure codes. 
Our method could be used for other types of condensed matter problems requiring integration over arbitrary convex VP.
Here, we implemented the general method in an site-centered, electronic-structure code and calculated formation enthalpies for FePd, yielding good agreement with experiment.
The radii to set the Voronoi/Delauney tessellation weights is obtained from a physics-based definition, i.e., the saddle-points in the total electron density. 

\vspace{0.2cm}
\emph{Acknowledgements:} Our work was supported by the U.S. Department of Energy BES/Materials Science and Engineering Division from contracts with Illinois (DEFG02-03ER46026), Ames Laboratory (DE-AC02-07CH11358), operated by Iowa State University, and Lawrence Livermore National Laboratory (subcontract B573247). DDJ and SNK acknowledge support from NSF (DMR-07-05089) and, recently, the Center for Defect Physics, Energy Frontier Research Center.  We thank Wan-Tsang Wang (NSYSU, Taiwan, and advisor Y.C. Chang) for help in algorithm testing while in Illinois.

\appendix
\section{Exact Solution for Cubic structures} \label{app1}
For volume conservation the VP and inscribed volume are known analytically, the interstitial integral is
\be
\Omega^{IS}_{\text{exact}}=\Omega^{VP}-\frac{4\pi}{3} [R_{MT}^{(s)}]^3,   
\ee
where $(s)$ indicates the lattice type (SC, FCC or BCC).
$R_{MT}^{(s)}$ is the inscribed sphere or (muffin-tin) radius of the lattice $s$. 
For charge conservation, it is straightforward to show that 
\be
Q^{IS}_{\text{exact}} = \frac{A^{(s)} \pi}{[T^{(s)}]^3} \left[\rule{0mm}{4mm} \sin\alpha^{(s)}-\alpha^{(s)} \cos\alpha^{(s)}  \right],
\ee
where $\alpha^{(s)} = T^{(s)} R_{MT}^{(s)}$, $T^{(s)}\equiv \vert{\bf T}_{n}\vert$. 
$A^{(s)}$ is a normalization constant whose value is $24$, $32$ and  $48$ for the SC, FCC, and BCC lattice, respectively.
Finally, the exact expression for $\rho V$ integral~\eqref{eq:rhoV} for the van Morgan charge and potential   for any general lattice can be simplified for the cubic lattices (SC, FCC, and BCC) as
\begin{eqnarray}
[\rho V]^{IS}_{\text{exact}}&=& \frac{4 \pi K^{(s)} [\Omega^{IS}]^{(s)}} {[T^{(s)}]^2} - \\
 && \frac{4 \pi}{[T^{(s)}]^3}  \sum_{i=1}^{p^{(s)}} f_{i}^{(s)}   \left[  \frac{sin~\gamma_{i}^{(s)} - \gamma_{i}^{(s)} cos~\gamma_{i}^{(s)}}{ [\beta_{i}^{(s)}]^3} \right],\nonumber
\end{eqnarray}
where $K^{(s)}= K =6,8,12$ and  $p^{(s)}=2,3,4$ for the three lattices, respectively, and $\gamma_{i}^{(s)} = \beta_{i}^{(s)} \alpha^{(s)}$.

{\par}For the above integral expressions, the coefficients  $f_i$ and $\beta_i$ for the SC, FCC, and BCC lattices are
\begin{eqnarray}       
\textrm{for SC,}\quad~f_1&=&\frac{f_2}{K-2}=K \ \ ; \  \beta_1=\sqrt{2}\beta_2=2.\nonumber 
\end{eqnarray}
\begin{eqnarray}       
\textrm{for FCC,}\quad~f_1&=&\frac{2}{K-2}f_2=\frac{2}{K-2}f_3=K ; \quad\quad \nonumber\\ 
\beta_1  &=& \sqrt{2}\beta_2 =\sqrt{\frac{3}{2}}\beta_3 = 2.\nonumber 
\end{eqnarray}
\begin{eqnarray}       
\textrm{for BCC,}\quad~f_1&=& \frac{2}{K-4}f_2=\frac{2}{K-4}f_3=\frac{1}{2}f_4 = K ; \nonumber\\
\beta_1 &=& \frac{2}{\sqrt{3}}\beta_2 = 2 \beta_3 = \sqrt{2}\beta_4 = 2.\nonumber  
\end{eqnarray}



\begin{thebibliography}{113}
\expandafter\ifx\csname natexlab\endcsname\relax\def\natexlab#1{#1}\fi
\expandafter\ifx\csname bibnamefont\endcsname\relax
  \def\bibnamefont#1{#1}\fi
\expandafter\ifx\csname bibfnamefont\endcsname\relax
  \def\bibfnamefont#1{#1}\fi
\expandafter\ifx\csname citenamefont\endcsname\relax
  \def\citenamefont#1{#1}\fi
\expandafter\ifx\csname url\endcsname\relax
  \def\url#1{\texttt{#1}}\fi
\expandafter\ifx\csname urlprefix\endcsname\relax\def\urlprefix{URL }\fi
\providecommand{\bibinfo}[2]{#2}
\providecommand{\eprint}[2][]{\url{#2}}

\bibitem{Voronoi08} G. F. Voronoi, J. Reine Angew. Math {\bf 134}, 198 (1908).

\bibitem{Ellis70} D. E. Ellis and G. S. Painter, \PR\ B {\bf 2}, 2887 (1970).

\bibitem{Boerr88} P. M. Boerrigter, G. Te. Velde, and J. E. Baerends, Int. J. Quantum Chem.\ {\bf 33}, 87 (1988).

\bibitem{Averil89} F. W. Averill and G. S. Painter, \PR\ B {\bf 39}, 8115 (1989).

\bibitem{Velde92} G. Te. Velde and E. J. Baerends, J. Comput. Phys. {\bf 99}, 84-98 (1992), and refs. therein.

\bibitem{Finocchiaro98} Daniele Finocchiaro, Marco Pellegrini, and Paolo Bientinesi, J. Comput. Phys. {\bf 146}, 707-725 (1998).



\bibitem{EMTO}  L. Vitos, I. A. Abrikosov, and B. Johansson, Phys. Rev. Lett. {\bf 87}, 156401 (2001).

\bibitem{Gonis1991} A. Gonis, Erik C. Sowa, and P. A. Sterne, Phys. Rev. Lett. {\bf 66}, 2207 (1991).

\bibitem{LSMS} Yang Wang, G. M. Stocks, W. A. Shelton, D. M. C. Nicholson, Z. Szotek, and W. M. Temmerman
Phys. Rev. Lett. {\bf 75}, 2867 (1995).

\bibitem{FLAPW} E. Wimmer, H. Krakauer, M. Weinert, and A. J. Freeman \PR\  B~{\bf 24}, 864 (1981); D. J. Singh and L. Nordstrom, {\it Planewaves, Pseudopotentials and LAPW method}, 2nd ed. (Springer, Berlin, 2006).

\bibitem{Tanemura-83} M. Tanemura, T. Ogawa, and N. Ogita, J. Comput. Phys. {\bf 51}, 191 (1983).

\bibitem{Isoparametric} T. J. R. Hughes,{\it The Finite Element Method}, Dover Publications, Inc., Mineola, N.Y. 11501  (2000).

\bibitem{Gellatly81} B.J. Gellatly and J.L. Finney, J. Non-Cryst. Solids\ {\bf 50}, 313 (1981).

\bibitem{Aurenhammer} F. Aurenhammer, Siam J. of Computing {\bf 16}, 78 (1987).

\bibitem{Alam09} Aftab Alam and D. D. Johnson, \PR\  B~{\bf 80}, 125123 (2009).

\bibitem{Morgan77} J. van W. Morgan, \JPC\ {\bf 10}, 1181 (1977).

\bibitem{MECCA} D.~D. Johnson, A. Alam, and A.~V. Smirnov, {\it MECCA: Multiple-scattering Electronic-structure Calculations for Complex Alloys (KKR-CPA Program, ver. 1.9)} (University of Illinois, Illinois, 2008).

\bibitem{Goede97} A. Goede, R. Preissner, and C. Fr\"{o}mmel, J. Comput. Chem. {\bf 18}, 1113 (1997).

\bibitem{Sastry97} Srikanth Sastry, David S. Corti, Pablo G. Debenedetti, and F. H. Stillinger, \PR\ E~{\bf 56}, 5524 (1997).

\bibitem{Richard74} F. M. Richards, J. Mol. Biol.\ {\bf 82}, 1 (1974).

\bibitem{Medvedev-Gerstein} N. N. Medvedev, Dokl. Akad. Nauk\ {\bf 337}, 776 (1994); Phys. Dokl.\ {\bf 337}, 157 (1994); M. Gerstein, J. Tsay, and M. Levitt, J. Mol. Biol.\ {\bf 249}, 955 (1995).

\bibitem{ASW-ref} V. Eyert, {\it The Augmented Spherical Wave Method: A 
Comprehensive Treatment, Lecture notes in Physics}, Vol. 719, (Springer, Berlin, 2007).

\bibitem{Xiao} H. Xiao, Z. Gimbutas, Comp. and Math. with Applications {\bf 59}, 663-676 (2001).

\bibitem{Hasel-Conroy} C. B. Haselgrove, Math. Comp. {\bf 15}, 323 (1961); H. Conroy, J. Chem. Phys.\ {\bf 47}, 5307 (1967).

\bibitem{Becke88} A. D. Becke, J. Chem. Phys.\ {\bf 88}, 2547 (1988).

\bibitem{REF-GaussProduct} A. H. Stroud and D. Secrest, {\it Gaussian Quadrature
 Formulas}, Englewood Cliffs, NJ: Prentice-Hall (1966).

\bibitem{Slater64-Clement67} J. C. Slater, J. Chem. Phys.\ {\bf 41}, 3199 (1964); E. Clementi, D. L. Raimondi, and Q. P. Reinhardt, J. Chem. Phys. {\bf 47}, 1300 (1967).

\bibitem{Alam10} Aftab Alam, Brent Kraczek and D. D. Johnson, \PR\ B~{\bf 82}, 024435 (2010).

\bibitem{Bernal} Javier Bernal, {\it FORTRAN codes for Voronoi tessellation and Delauney triangulations}, NIST Math webpages, \urlprefix\url{http://math.nist.gov/~JBernal/JBernal_Sft.html}.

\bibitem{Johnson-fastKKR} D. D. Johnson, F. J. Pinski, and G. M. Stocks, Phys. Rev. B~{\bf 30}, 5508 (1984)

\bibitem{vonBH} U. von Barth and L. Hedin, \JPC\ {\bf 5}, 1629 (1972).

\bibitem{Monkhorst} H. J. Monkhorst and J.~D. Pack, \PR\ B~{\bf 13}, 5188 (1976).

\bibitem{KKR-CPA-Johnson} D. D. Johnson, D. M. Nicholson, F. J. Pinski, B. L. Gyorffy, and G. M. Stocks, Phys. Rev. Lett. ~{\bf 56}, 2088 (1986).

\bibitem{scr-cpa} D. D. Johnson and F.~J. Pinski, \PR\ B~{\bf 48}, 11553 (1993).

\bibitem{Gonis-Butler} A. Gonis and W. ~H. Butler, {\it Multiple Scattering Theory} (New York: Springer) (2000).

\bibitem{wang94} N. Stefanou, H. Akai, and R. Zeller, Comput. Phys. Commun.~{\bf 60}, 231 (1990); Y. Wang, G. M. Stocks, and J. S. Faulkner, \PR\ B~{\bf 49}, 5028 (1994).

\bibitem{Hultgren} R. Hultgren, P. D. Desai, D. T. Hawkins, M. Gleiser, and K.K. Kelley, {\it {Selected values of Thermodynamic properties of Binary Alloys} } (American Society for Metals, Metals Park, Ohio, 1973).

\bibitem{Wang2004} L. Wang, Z. Fan, A.~G. Roy, and D.~E. Laughlin, J. Appl. Phys. ~{\bf 95}, 7483 (2004).

\end{thebibliography}

\end{document}